# MgB$_2$ Bulk Superconducting Magnet

A. Yamamoto[1,3], A. Ishihara[2], M. Tomita[2], J. Shimoyama[1] and K. Kishio[1]

[1] Department of Applied Chemistry, University of Tokyo,
7-3-1 Hongo, Bunkyo-ku, Tokyo 113-8656, Japan
[2] Materials Technology Division, Japan Railway Technical Research Institute,
2-8-38 Hikari-cho, Kokubunji-shi, Tokyo 185-8540, Japan
[3] JST, PRESTO,
4-1-8 Honcho Kawaguchi, Saitama 332-0012, Japan

Various potential applications for the bulk superconducting magnet have emerged owing to recent progresses in superconducting bulk materials with high critical current density and mechanical strength [1] and performance of cooling systems which much easily provide low temperature environments. MgB$_2$ with $T_c \sim 40$ K has several attractive natures for the bulk superconducting magnet, such as weak-link-free homogeneous current flow on a bulk scale [2], great flexibility in designing magnet shape, low cost materials and light weight (1-2.6 g/cm$^3$). In the present study we have prepared a variety of disk-shaped MgB$_2$ bulk superconducting magnets (10, 20, 30 mm in diameter, 10, 20 mm in thickness) using the *in-situ* technique from Mg and B powders and evaluated the superconducting properties and trapped magnetic field properties.

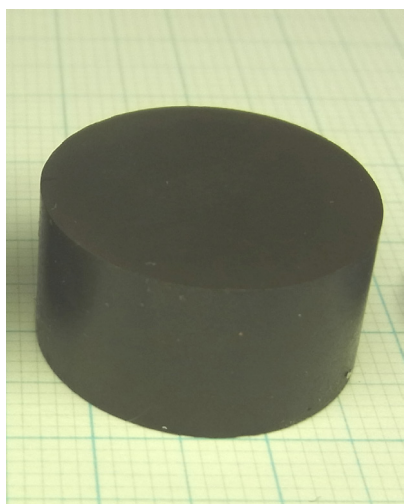

Fig. 1. Photograph of a disk-shaped MgB$_2$ bulk (20 mm in diameter, 10 mm in thickness).

Figure 1 shows a photograph of a disk-shaped MgB$_2$ bulk. The bulk showed homogeneous surface texture in macroscopic scale. Indeed we did not observe any cracks or domain structures, which disturb bulk current flow and distribution of trapped magnetic field [3]. Electromagnetic measurements revealed the obtained MgB$_2$ bulks have high critical current density of $>10^5$ A/cm$^2$ at 15-30 K. For the trapped magnetic field measurements, the bulk samples were cooled down by a cryocooler and magnetized by a superconducting magnet under the field-cooling condition. Trapped magnetic field of the bulk magnets was measured by a transversal cryogenic hall sensor. Figure 2 shows trapped magnetic field as a function of cold-stage temperature. MgB$_2$ bulk disks with 20 mm$\phi$ and 30 mm$\phi$ showed 1.8 T and 2.25 T at 15 K, respectively, at the center of bulk surface. Then we sandwiched a hall sensor between two bulks and measured a trapped magnetic field of the disk pairs in order to measure a trapped field comparable to that of inside of the disk. The disk pair with 20 mm$\phi$ showed higher trapped magnetic field of 2.8 T at 15 K, which is more than 50% higher than that of the surface of the single disk. We observed a further higher trapped magnetic field of above 3 T at 17.5 K for the 30 mm$\phi$ disk pair, promising for new compact bulk superconducting magnets operating at 15-30 K with a cryocooler.

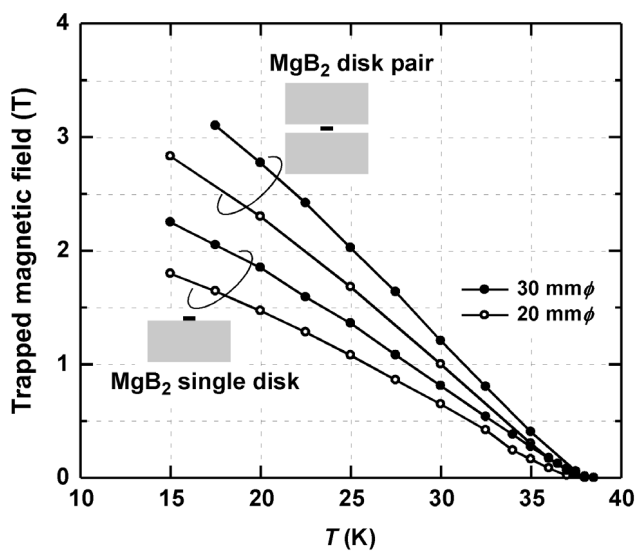

Fig. 2. Trapped magnetic field as a function of cold-stage temperature for MgB$_2$ bulk disks with 20 or 30 mm$\phi \times$ 10 mm$^t$ and disk pairs with 20 or 30 mm$\phi \times$ 20 mm$^t$.